\documentclass[review]{elsarticle}
 \usepackage{lineno,hyperref}
 \modulolinenumbers[5]

\journal{Journal of New Astronomy}

\begin{document}

\begin{frontmatter}

\title{Inversion of Zeeman polarization  for solar magnetic field  diagnostics}

 
 \author{M. Derouich$^{1,2}$}
 \address{$^{1}$Astronomy  Department, Faculty of Sciences, King Abdulaziz University, 21589 Jeddah, Saudi Arabia\\
 $^{2}$Sousse University,  ESSTHS, Lamine Abbassi street, 4011 H. Sousse, Tunisia}


 \ead{derouichmoncef@gmail.com}


\begin{abstract}
The topic of magnetic field diagnostics with the Zeeman effect is currently vividly discussed. 
There are some testable  inversion codes available to the spectropolarimetry community and their application allowed for a    better understanding of the magnetism of the solar atmosphere.   In this context, we propose an inversion technique associated with a new numerical code. The inversion procedure is promising and   
particularly successful for interpreting the Stokes profiles in quick and sufficiently precise way.    In our inversion, we fit a part of each Stokes  profile  around a target wavelength,  and then determine the magnetic field as a function of the wavelength  which is
equivalent to get the magnetic field as a function of the   height  of line  formation.

To test  the performance of the new numerical code, we employed ``hare and hound'' approach by comparing an exact solution (called input) with the solution obtained by the code (called output).  The precision of the code is  also  checked by comparing our results to the ones obtained with the HAO MERLIN code. The inversion code has been applied to synthetic Stokes profiles of the Na D$_{1}$ line available in the literature. We   investigated the limitations in recovering
the input field in case of noisy data.  As an application, we  applied our inversion code to the polarization profiles of the Fe {\sc i} $\lambda$ 6302.5 \AA\   observed at IRSOL in Locarno.
\end{abstract}


\end{frontmatter}

 \section{Introduction}

 In the quiet Sun, where the magnetic field is weak, the Zeeman splitting can be smaller than the Doppler width of the spectral lines; in these conditions the Zeeman effect cannot be observed and one must use the Hanle effect technique to obtain the solar magnetic field  (e.g. Stenflo 1982;  Landi Degl'Innocenti  1983; Sahal-Br\'echot et al 1986; Trujillo Bueno et al. 2004; Derouich et al. 2006; Faurobert et al. 2009; Derouich et al. 2010).  However, in the active regions  where the magnetic field is sufficiently strong, the Zeeman effect produces a measurable splitting of the atomic levels and a subsequent polarization of the emitted light  (e.g. Harvey et al. 1972;  Jefferies et al.   1989;  Socas-Navarro et al. 2000; Asensio Ramos et al. 2012).

Rigorous interpretation of the Zeeman effect on the  spectral  polarization   can be a crucial source of information about the  Sun's   magnetic field. It is necessary to apply suitable theoretical and numerical methods  to extract  the physical information from spectro-polarimetric solar observations. To this aim, from   the beginning of the 70's of the 20th century,    vigorous theoretical and numerical  efforts have been made to develop  non-linear inversion codes that  are able to reliably derive information about the   magnetic properties of the solar plasma.  The present work is a new contribution to  these efforts. Our aim is to illustrate a new inversion procedure 
in order to search  for new possibilities to determine the structure and 
distribution of the solar magnetic field.   The inversion technique  uses the equations
established by Landi Degl'Innocenti  and Landi Degl'Innocenti (1972),  and later by Jefferies et al.  (1989) and Stenflo  (1994). 


   In  this work, we focus on the accuracy of the fitting
 algorithms of the Stokes profiles to ensure maximum performance of the  inversion technique.
In fact, for the familiar  Zeeman effect, the responsible physical 
mechanisms are  typically already well understood\footnote{This is in contrast of   the Hanle effect 
for which  a variety of less familiar physical mechanisms are still not well understood.}. 
Therefore, the advances in the inversions codes, which are based on the Zeeman effect, should be   mainly 
concentrated on how to fit the Stokes profiles in order to deduce the   magnetic field. It is the intention of this paper 
 to find out the best fitting strategy. Careful fitting   of the Stokes profiles allows to reduce significantly the error 
 bar on the determination of the magnetic field. This work proposed  useful  methods  to achieve a proper fit of the Q, U, V, and 
 I-profiles.   
 
In order to validate our numerical code, through a controlled and pragmatic strategy,    we adopt  ``hare \& hound'' approach consisting of the following steps:

(1) We make use of  synthetic Stokes profiles of the Na I $\lambda$ 5896 \AA\ line available in the literature (Uitenbroek (2001, 2003,   2011), Leka et al. 2012). The  magnetic  maps  which served to generate   these Stokes profiles, with the aid of 3D NLTE radiative transfer  models, are also available and are called {\it input}. 

(2)  We   analyze the synthetic Stokes profiles using our inversion code, and attempt to retrieve the magnetic maps (called {\it output})

(3) We compare  the exact solution (input) with the solution provided by  the inversion code (output).

(4) We analyze the dependence of the results on the signal-to-noise ratio. Different levels of expected noise 
  are then simulated to evaluate their impact on the precision of our results.
This  allows us to determine the needed polarimetric sensitivity and estimate the error bars in the determination of the solar magnetic field. 

In addition, the precision of the code is  also  checked by comparing our results to the ones obtained with the HAO MERLIN code.  The comparison is based on the   level 1 and level 2 data available online at the Community Spectro-polarimtetric Analysis Center.\footnote{CSAC; http://www.csac.hao.ucar.edu/}
Finally, using our numerical code we   interpret    Fe {\sc i} $\lambda$ 6302.5 \AA\   observed at IRSOL in Locarno.

 \section{Inversion method}
 The inversion formalism  is based  on the equations
 obtained by Landi Degl'Innocenti  and Landi Degl'Innocenti (1972),  and later by Jefferies et al.  (1989) and Stenflo  (1994). According to these equations, the  Stokes parameters are related in a simple way to the  magnetic field vector   under   the weak field approximation.  The domain of validity of these equations  was reviewed and presented in Table  
 9.1 of a  monograph by Landi Degl'Innocenti  \&  Landolfi (2004). The inversion  code  allows us to deduce the   magnetic field by fitting theoretical profiles    to the observed ones.

  \subsection{Fitting of the $I$-profiles}

\begin{figure}
\begin{center}
\includegraphics[width=14 cm]{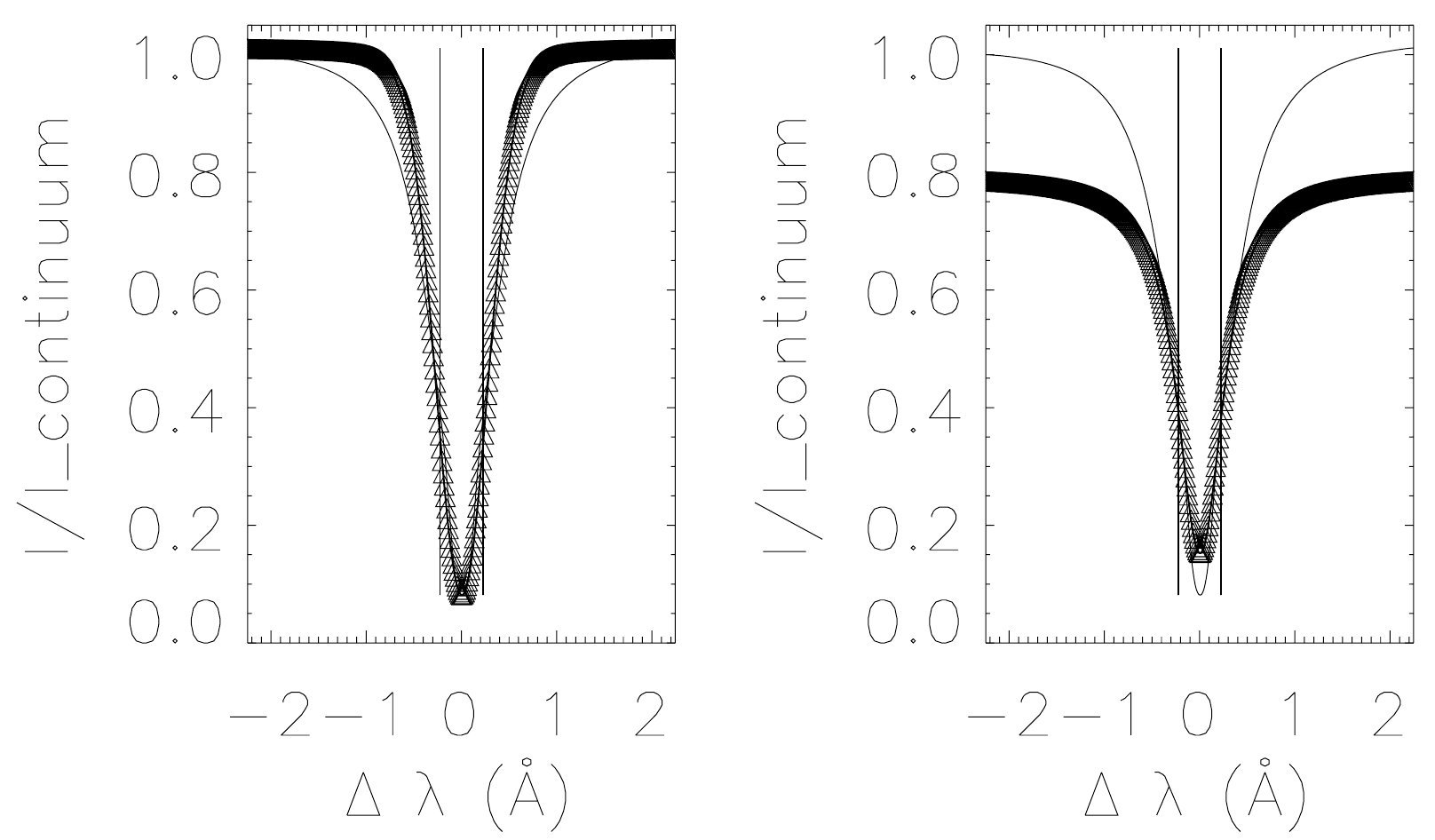}
\end{center}
\caption{ \textsf{
A  plot representing a Voigt function fit of the intensity profile of an umbra pixel. This fit is performed as follows. The  open triangles ($\triangle$)  represent  the theoretical profile  resulted from the fitting procedure. Left:     the weighting is taken in a manner      to  improve the determination   of the    continuum
 intensity, the   line strength and the line center.   The vertical line  represents the place of the target wavelength	   $\delta\lambda_B$= 0.224520 \AA. 
 Right:  the weighting  is changed in a way to obtain the best Voigt-fitting  around the  
  target wavelengths where the magnetic field will be 
determined.   }}
\label{perturbed}
\end{figure}

We adopted, the well known Voigt line shape  to fit the intensity profiles.   Voigt function depends on five parameters: the continuum of the profile,  the line strength, 
the damping parameter, the Doppler width, and the value of the line center. The fitting consists of the determination of these five parameters.
Our fitting strategy   is divided into three steps. The  first step consists of obtaining a first guess. In this first step we use an uniform weighting  to all points in the profile. Using the results of the initial analysis, as a second step,  we change the    weights of some points in the sense of   improving the determination   of the    continuum
 intensity, the   line strength and the line center.  The inversion method developed in this work permits the magnetic field  determination  in a given  target wavelength $\delta\lambda_B$.    Therefore, in the third step  only a part of the intensity profile surrounding $\delta\lambda_B$ is well fitted. With this aim, the choice of weighting  is done in such a manner as to obtain the best Voigt-fitting  around the   target wavelengths where the magnetic field is determined.  The  third step  permits especially the determination of 
 the Doppler width and the Doppler damping. Figure \ref{perturbed} represents an example of the fitting of an intensity profile where the second and the third steps are illustrated.

Now one uses the results of the fitting of the $I$-profile   to obtain the Voigt function  $H(a,v)$  and the Faraday-Voigt function  $F(a,v)$;  $H(a,v)$ is   associated to the absorption coefficient which is  proportional to the imaginary part of the complex refractive index and 
  $F(a,v)$ is  associated to   the real part of the complex refractive index.    Note that   $F(a,v)$ indicates the   measurable  consequences of the  changes of  the phase velocity of the wave in the atmosphere.    Here $a$ is the Voigt parameter and $v = \delta\lambda_B$/$\delta\lambda_D$, where the target  wavelength 
$\delta\lambda_B$ 
 represents the shift in wavelengths due to the   Zeeman
effect of the magnetic field and $\delta\lambda_D$  is the Doppler width of the intensity profile.
  In addition, one can  determine  the  values of     $H'(a,v)=\frac{\partial H}{\partial v}$, $H''(a,v)=\frac{\partial^2 H}{\partial^2 v}$ and $\frac{\partial I}{\partial \lambda}$
 which gives the variation 
of the intensity $I$ (in arbitrary units) for a small variation   $d\lambda$ (in  Angstr\"om) along 
the $I$-profile. Considering that  $S$ designates the line strength, in the Voigt fitting case, one can show that, 
\begin{eqnarray}\label{eq_didl}
\frac{\partial I}{\partial \lambda}= -2  S  \times     [   2 \quad a \quad F(a,v)-v \quad H(a,v)]/\delta\lambda_{D}
\end{eqnarray}

  \subsection{Fitting of the  polarization profiles}
\begin{figure}
\begin{center}
\includegraphics[width=12 cm]{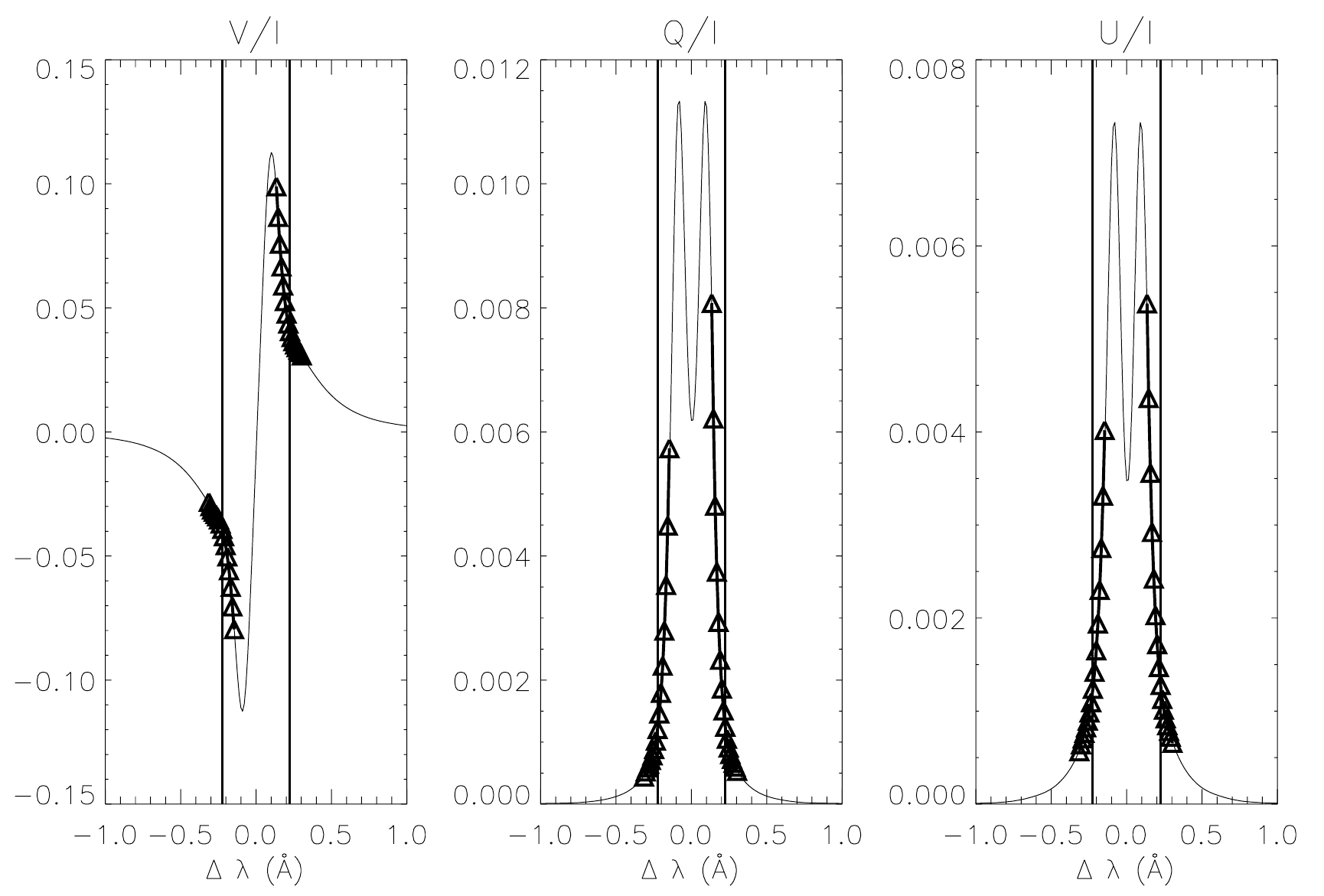}
\end{center}
\caption{ \textsf{ Fit of  Q, U and V profiles of an umbra pixel:  SVDC and SVSOL methods 
 applied to Non-LTE synthetic Na I $\lambda$ 5896 \AA\ line. The  open triangles ($\triangle$)  represent  the theoretical profile  resulted from the fitting procedure.  The vertical lines represent the place of the  target wavelength	  $\delta\lambda_B$= 0.224520 \AA.     }}
\label{Stokes-Fitting}
\end{figure}

Once the intensity profile is well-fitted, we  fit the Stokes parameters $Q$, $U$, and $V$.     The fitting of Stokes $Q$, $U$ and $V$ is obtained   with the 
Singular Value DeComposition (SVDC) and Singular Value SOLve (SVSOL) routines.    
The SVDC   expands   the original   
data that we are trying to fit  in a  4-dimensions basis. We checked other possible dimensions and we found that   a  4-dimensions basis is  more appropriate. After that,  the  SVSOL uses the results of the expansion  generated by SVDC in order to fit the polarization profiles, i.e. obtaining 
the solution which corresponds to the minimization of the $\chi^2$. An example of the result of the fitting of the  polarization profiles is given in the Figure \ref{Stokes-Fitting}.

It is worth noticing that the fitting of the Stokes parameters  is performed after convolution of the  original spectra.  The convolution   is typically needed 
 to smooth the profile and   to decrease the effect of the noise.  Generally, one can perform the convolution  with a rectangle or  kernel functions,   or with an instrumental profile. 
  \subsection{From the best fit profiles to the magnetic vector}
\begin{figure}
\begin{center}
\includegraphics[width=10 cm]{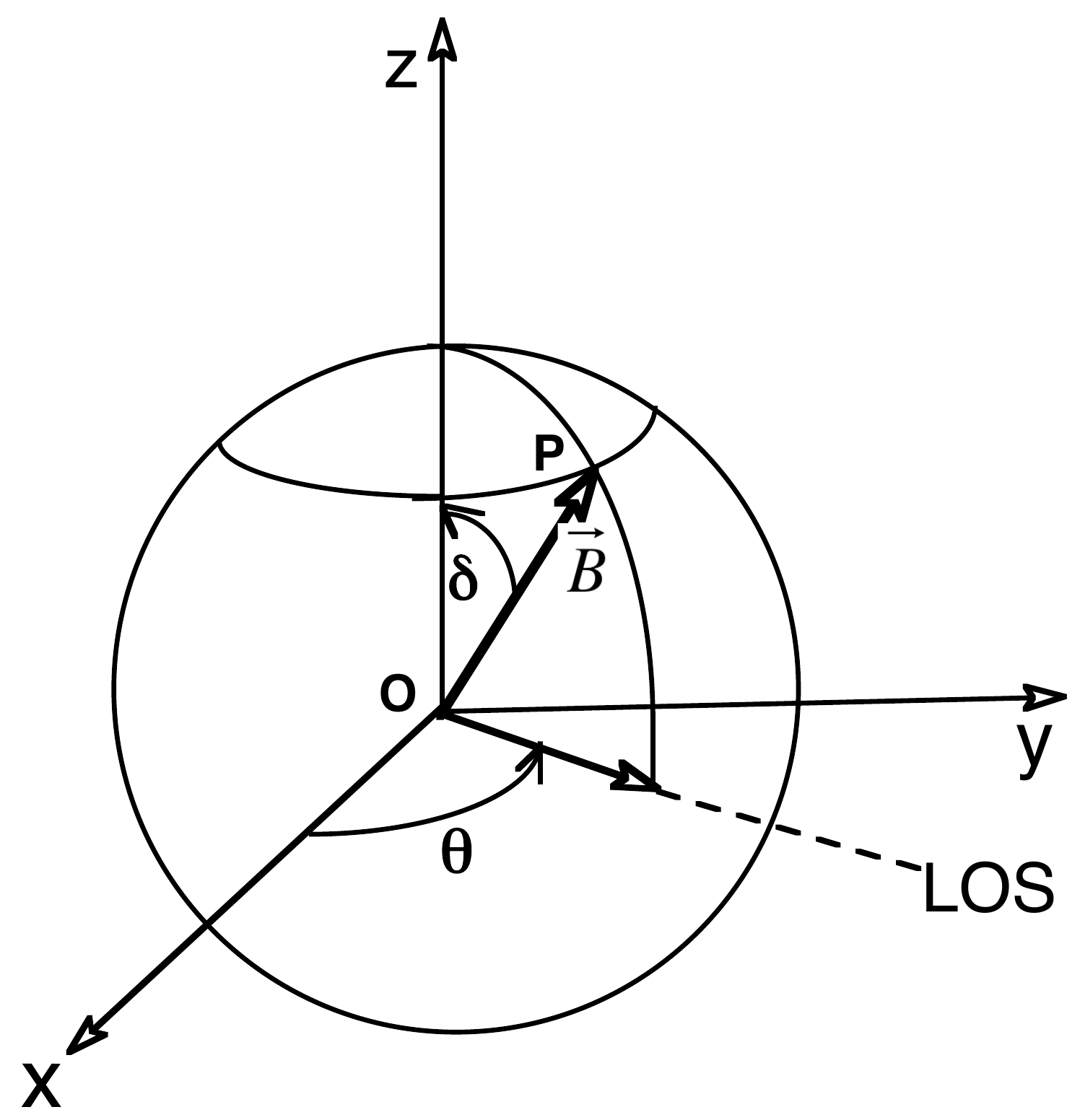}
\end{center}
\caption{ \textsf{An illustration of the   Cartesian and spherical coordinates of the magnetic field.}}
\label{coordinates}
\end{figure}

 The   magnetic vector can be determined  in the    Cartesian coordinates    ($B_x,B_y,B_z$) or   in the spherical coordinates ($B$, $\theta$, $\delta$).   The representation of these coordinates is illustrated in the Figure \ref{coordinates}.
Geometric relations between  
the  Cartesian components of the magnetic field ($B_x,B_y,B_z$) and its spherical coordinates ($B$, $\theta$, $\delta$) are:
\begin{eqnarray}  \label{PaperKS_eq_1}
B_x & = &     B  \sin\delta  \cos\theta    \nonumber \\
B_y & = &     B \sin\delta \sin\theta       \\
B_z & = &     B \cos\delta      \nonumber
\end{eqnarray} 
 The longitudinal part of the magnetic field is  $\vec{B}_{\small LOS}$=$\vec{B}_{z}$  and the transverse 
 part is  $\vec{B}_{\small Trans}$=$\vec{B}_{x}$+$\vec{B}_{y}$. The angles $\delta$ and   $\theta$ satisfy the following equations:
\begin{eqnarray} \label{PaperKS_eq_2}
\theta & = &   \tan^{-1}  (\frac{B_y}{B_x})  \nonumber \\
  - \pi   \le \theta & \le &   \pi 
\end{eqnarray} 
and,
\begin{eqnarray}  \label{PaperKS_eq_3}
\delta  & = &     \cos^{-1}  (\frac{B_z}{B})  \nonumber \\
-\frac{\pi}{2}  \le \delta & \le &    \frac{\pi}{2}   
\end{eqnarray} 
 
 After solving the radiative transfer equations, one could demonstrate that (e.g. Landi Degl'Innocenti  and Landi Degl'Innocenti (1972)   and  Jefferies et al.  (1989)):
\begin{eqnarray}\label{eq_BLOS}
B_{\small LOS}(Gauss)=\frac{-2.142 \times 10^{12}}{g \times \lambda_0^2} \times  \frac{V}{(\frac{\partial I}{\partial \lambda})},
\end{eqnarray}
 and  the transverse  magnetic field:
\begin{eqnarray}\label{eq_BTrans}
 |B_{\small Trans}(Gauss)|  & = &   2 \times \frac{2.142 \times 10^{12}}{g \times \lambda_0^2} \\
&& \times 
\sqrt{\frac{|H'(a,v)|}{v \times |H''(a,v)|}   \times \sqrt{(Q^2+U^2)} 
\times |\frac{\delta\lambda_B}{(\frac{\partial I}{\partial \lambda})}|}  \nonumber
\end{eqnarray} 
and the inclination:
\begin{eqnarray}\label{eq_Azimut}
 \theta  = \frac{1}{2} \tan^{-1}(\frac{U}{Q})
\end{eqnarray} 
Equations (\ref{PaperKS_eq_3}, \ref{eq_BLOS}, \ref{eq_BTrans}, \ref{eq_Azimut}) are used through this paper to determine the magnetic maps.

Note that the equation (\ref{eq_BTrans})  gives only the absolute value of  $B_{\small Trans}$. It means that  the fundamental ambiguity is not solved.  Two field vectors that are symmetrical with respect to the line-of-sight (LOS) have the same polarimetric signature.

 \section{Results of the inversion	of the synthetic Stokes profiles of the  Na I $\lambda$ 5896 \AA\ line}
 Our objective is to determine the values of the Stokes spectra at the 
target wavelengths $\delta \lambda_B$ and to introduce them in the Equations 4--7 to obtain the magnetic vector. We do not fit the whole profile but we only fit 
the part of the profile defined by 
\begin{eqnarray}\label{eq_3_range}
 \delta \lambda = \delta \lambda_B    \pm \delta \lambda_{TOL}  
\end{eqnarray}
$\delta \lambda$ is the shift in wavelength	s from the line center and the   $\delta \lambda_{TOL}   $ is called    tolerance, which  is arbitrary. In principle the value of the magnetic  field vector does not depend on 
the choice of the tolerance value. However,   it should be noticed that for small values 
of tolerance the magnetic field jumps very quickly which means that for small values 
of tolerance   the inversion method is not stable numerically.    Typically, 30 mA $\le \delta \lambda_{TOL} \le$ 90 mA.

\begin{figure}
\begin{center}
\includegraphics[width=13 cm]{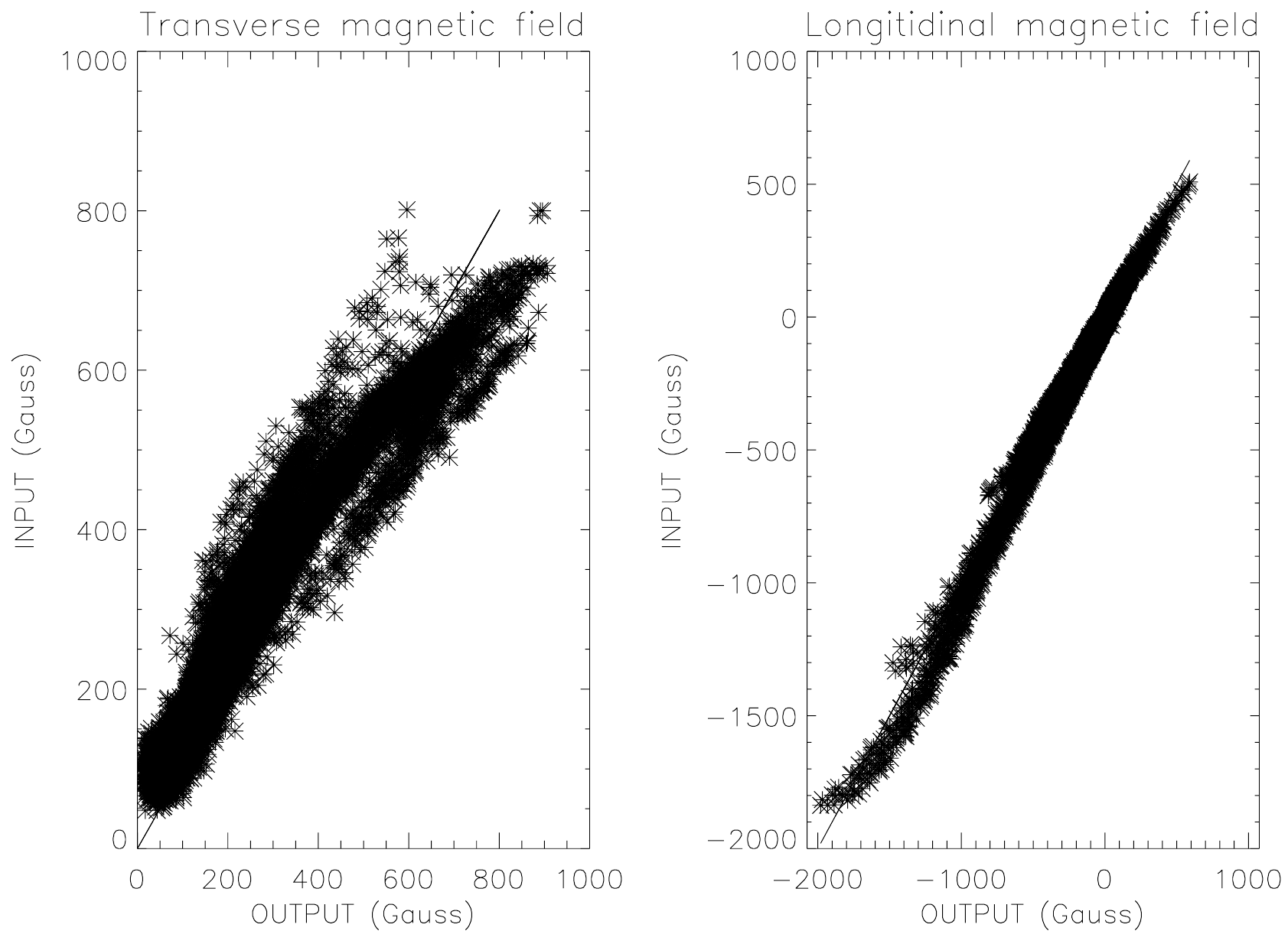}
\end{center}
\caption{ \textsf{
A  plot representing  the output and the input of the longitudinal and transverse magnetic fields. A noise level  =   $10^{-4}$ is added to the  theoretical Stokes profiles used in the inversion.   The output is showed  with stars  $   \ast  $.}}
\label{figureInputVSoutputLOS}
\end{figure}

 To test  the performance of our fitting method we employ ``hare \& hound'' approaches consisting of comparison between the exact solution (input) and the solution provided by  the inversion method (output).  Figure  \ref{figureInputVSoutputLOS} represents    the theoretical perfect correspondence (output=input)  and the result of the fit  (output vs input).   It is to be mentioned here  that the NLTE  approach and the atomic model used to generate the Stokes vector (Uitenbroek (2001, 2003,   2011), Leka et al. 2012) is different from the NLTE  approach and the atomic model adopted in our inversion method, which would make  some inevitable  differences between input and output magnetic fields even in zero-noise case.

 A noise level  =   $10^{-4}$   is added to a map of 142 $\times$ 128 pixels.  Each pixel contains    the synthetic  Stokes profiles.  We invert  the data  obtained   for  an observing angle $ \mu$=1 where, for the synthetic  Stokes profiles studied here, the circular polarization is clearly larger than the linear polarization (see  Leka et al. 2012). As a result of the inversion, we find that the averaged relative error in the case of the longitudinal magnetic field is    
 12\%.  For the transverse component, we found that 
 the relative error is   35\%. In the case of the inclination angle, the averaged relative error is   30\%.   
  The  relative error is smaller in the case of the  longitudinal magnetic 
  field due to its dependence   on the circular polarization $V$  which is sufficiently large. On the contrary, 
  the transverse field depends on the linear polarization which is   small and its determination is very sensitive to the noise.
The inclination depends on both the linear and the circular polarizations.    

Let us mention that the synthetic Stokes profiles contain  pixels in quiet Sun, plage, umbra and penumbra. Thus, one should take into account that, for example, magnetic fields in quiet Sun pixels with absolute values lower than 10 Gauss are not able to reproduce measurable Zeeman effect   and thus are not easily recovered in the output.  This could explain why the averaged error seems to be rather large especially in the case of the transverse field and the inclination.

In the case of the   azimuth, we found difficulties  in comparing the input with the output.  In order to understand the source of these difficulties, we decided to compare  the azimuth angle using the input magnetic field via the equation   $\theta_1   =       \tan^{-1}(\frac{B_y}{B_x})   $ to the azimuth angle that one obtains    from the equation 
  $ \theta_2  = \frac{1}{2} \tan^{-1}(\frac{U}{Q})$.  Interestingly, $U$ and $Q$ are the  synthetic  NLTE emergent Stokes parameters generated from the input magnetic field  $\vec{B}$($B_x$, $B_y$, $B_z$), i.e. $\theta_1$ and $\theta_2$ are  inferred from the input. Thus, in principle, $\theta_1$ and $\theta_2$ must be similar. However, we found them quite different. The possible explanation of this discrepancy is   that the components $B_x$ and $B_y$ are determined in a reference   different  from the reference in which Stokes parameters were calculated. 
 Consequently,   in the case of the   azimuth, we could not compare correctly the input with the output. 
 
 Note that only the Stokes parameters  $U$ and $Q$ are defined with respect to a given reference direction.  When the reference direction changes, $Q$ and $U$ tend to change into each other. 
 The Stokes $V$, the intensity $I$ and the complete linear polarization $\sqrt{U^2+Q^2}$  are invariant under rotation of the reference direction.
 
 \begin{figure}
\begin{center}
\includegraphics[width=7 cm]{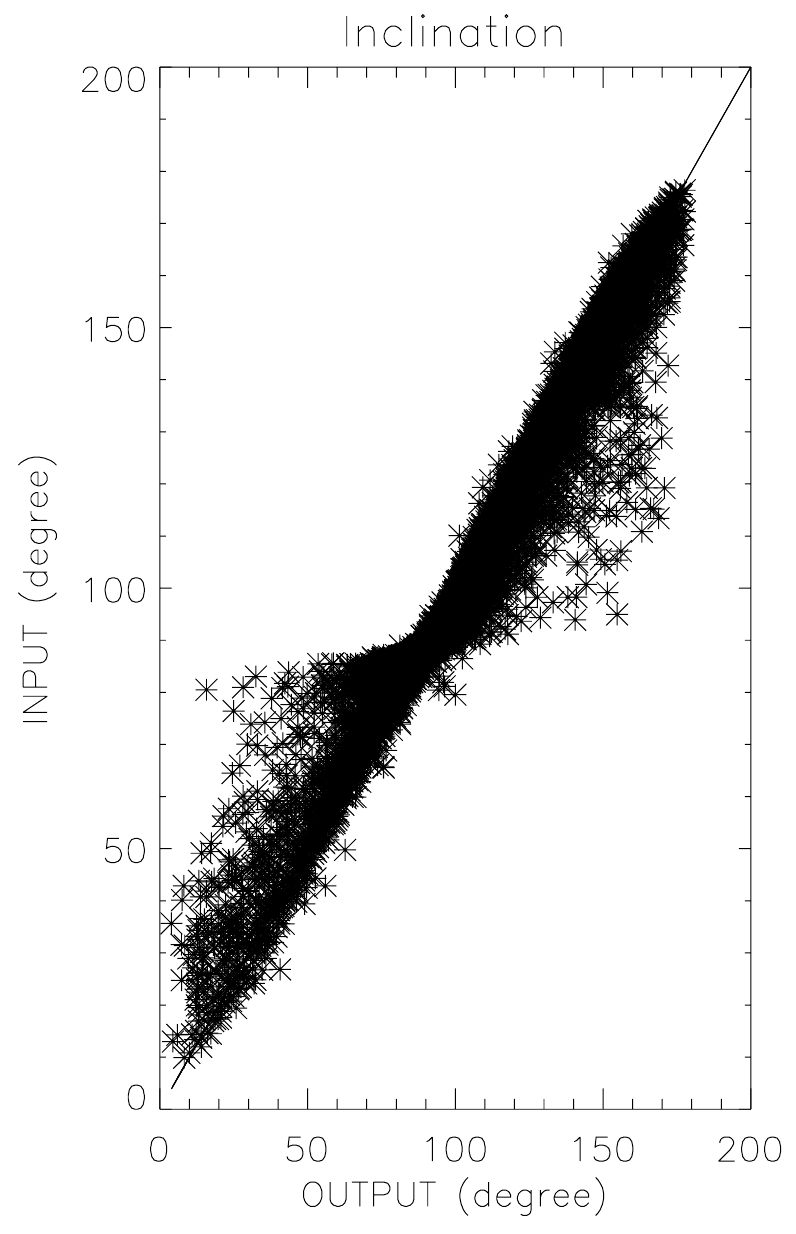}
\end{center}
\caption{ \textsf{
A  plot representing  the output and the input of the Inclination. A noise level  =   $10^{-4}$ is added to the  theoretical Stokes profiles used in the inversion. The output is showed  with stars  $   \ast  $.}}
\label{figureInputVSoutputINCLIN}
\end{figure}
 
 \section{Effect of the noise}
 In order to evaluate   the effect of the noise, we investigate the reliability of the fitting accuracy in case of noisy Stokes spectra by computing the standard deviation $\sigma$. Furthermore, we compare the magnetic field  obtained in a zero noise case to the one obtained in a noisy case.
  \subsection{Standard deviation $\sigma$ for polarization profiles fitting}
In our work we use numerically-generated Stokes profiles  from a known magnetic model (Uitenbroek (2001, 2003,   2011), Leka et al. 2012). 
These profiles are artificially polluted with   noise. At each noise level, one added poisson-distributed noise to an umbra pixel and 1000 noise realizations of the data were generated.  We compute the standard deviation   $\sigma$  which is given by :
\begin{eqnarray} 
\sigma= \sqrt{\frac{\sum (x_i-x_{exact})^2}{N-1}}
\end{eqnarray}
where, $i$ is an index to indicate a given  realization and N is the total number of  realizations (N=1000). The symbol $x_i$ represents  a Stokes parameter  $Q/I$, $U/I$, or $V/I$ of a given realization and
$x_{exact}$ is the exact value of the Stokes parameter at the umbra pixel. Our  fit  via the SVDC and SVSOL procedures was performed around  a target wavelength $ \delta \lambda_B$=0.134712 \AA   and the tolerance $ \delta \lambda_{TOL}$=45 m\AA.  
  Table  \ref{table_deviation}  shows   low standard deviations   $\sigma_{V}$,  $\sigma_U$ and $\sigma_Q$ which   means that the Stokes profiles are    well fitted and the results must be reliable.  
 \begin{table} 
\begin{tabular}{ l| c| c| c| c|c|c| r }
\hline
  $\sigma_{V}$ &0.00376968 &  0.00198307 & 0.000884621 & 0.000640550  &0.000548571&  0.000527108  &0.000514566
\\ 
\hline
 $\sigma_U$& 0.00404129  & 0.00193955 & 0.000788354 & 0.000412921  &0.000258710  &0.000213259 & 0.000175086 \\
\hline
 $\sigma_Q$& 0.00395388   &0.00202286  &0.000896636 & 0.000575089 & 0.000443220  &0.000431046  &0.000412398 \\
\hline
 noise   & 0.005 &0.0025 &  0.001 &  0.0005 &   0.0001 &  5 $\times$ 10$^{-5}$ &     0.00  \\
\hline
\end{tabular}
 \caption{$\sigma$ values  resulted from the SVDC and SVSOL  fitting. }
\label{table_deviation}
\end{table}
    \subsection{Magnetic field in a zero noise case vs noisy case}
\begin{figure}
\begin{center}
\includegraphics[width=7 cm]{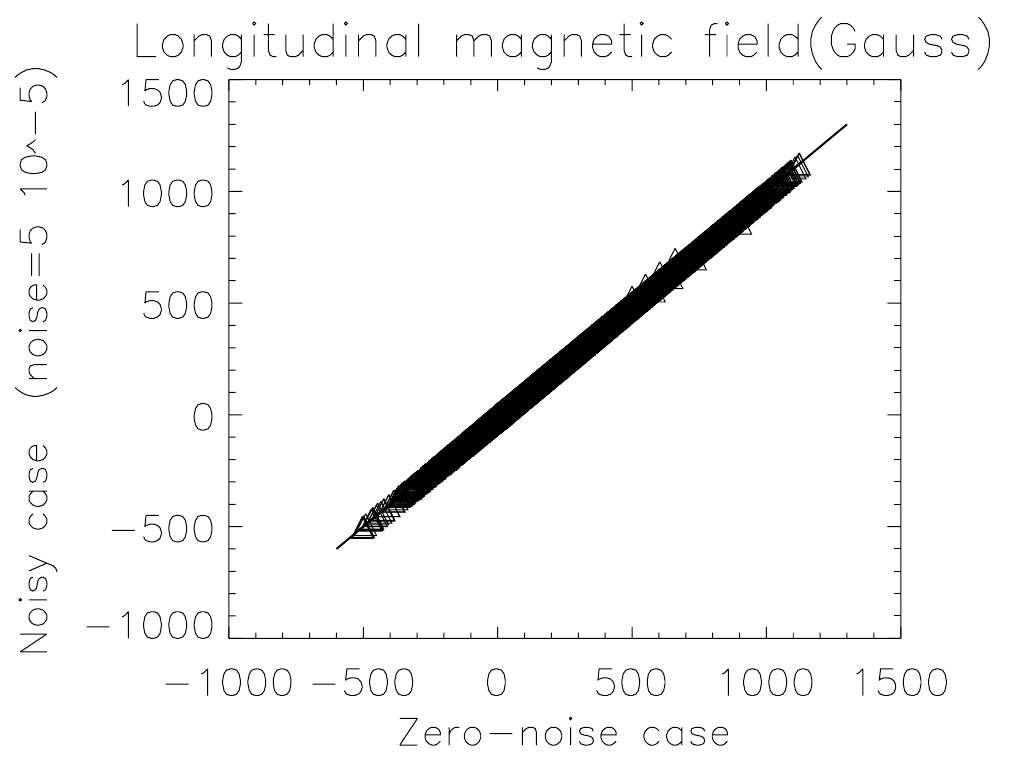}
\includegraphics[width=7 cm]{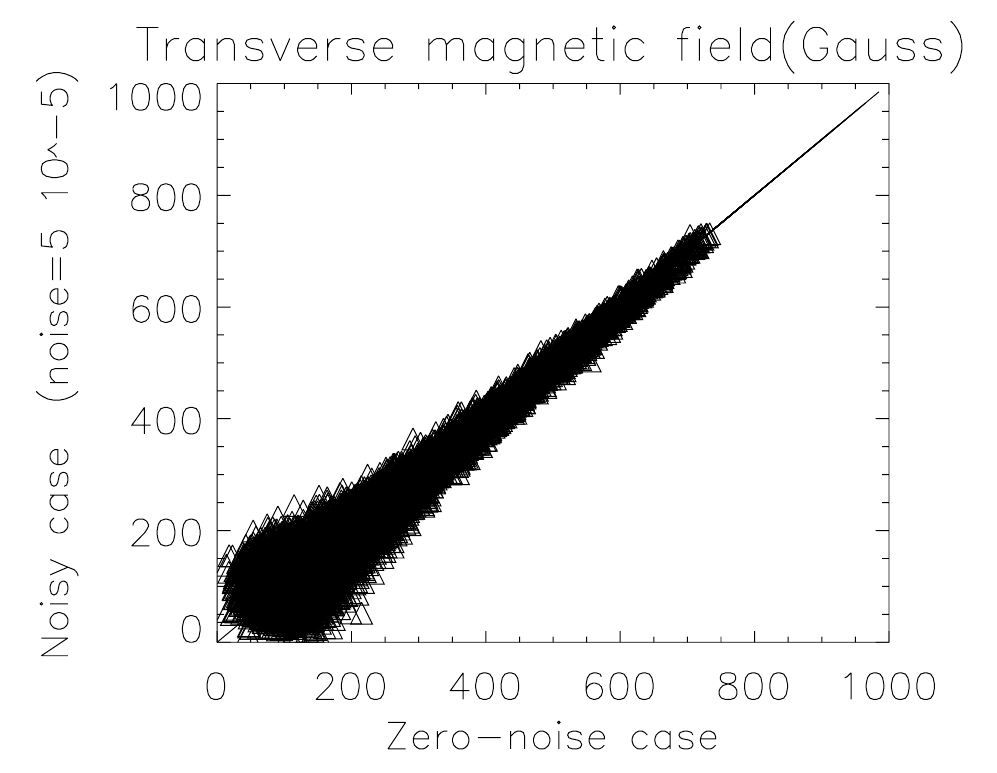}
\end{center}
\caption{ \textsf{
A  plot representing  the effect of a noise level= 5  $\times$10$^{-5}$.  The solid line shows the theoretical  exact solution   corresponding to the equation $y=x$ in the Cartesian coordinates. The  open triangles   ($\triangle$) represent   the magnetic field obtained in the  noisy case as a function of the magnetic field derived in the zero noise case. }}
\label{figureB_noise10_4}
\end{figure}

 As we mentioned previously, even in zero noise case,  some inevitable  differences between input and output magnetic fields occur.  In order to exclude the effects of other factors  and to evaluate exclusively the effect  of the noise, we compare a noisy output to a zero noise output. 
Therefore,  we compare  
the magnetic field derived in the case where the effect of the noise   is added to the one derived in the zero noise  case.   We start by  adding a low noise level=5 $\times$ 10$^{-5}$.
 Figure    \ref{figureB_noise10_4}    shows the   comparison between the exact solution corresponding to the equation of a straight line passing through the origin  and  the solution affected by the noise.
  It is clear from the Figure   \ref{figureB_noise10_4}   that the  effect  of the noise is   small and the accuracy for the determination of the magnetic field is good. 
This is especially the case for the longitudinal component where  the difference between the noisy and non-noisy cases is less than 5\%. In the case of the  transverse magnetic field, the difference can reach up to 
 25\%.

\begin{figure}
 \begin{center}
 \includegraphics[width=7.5 cm]{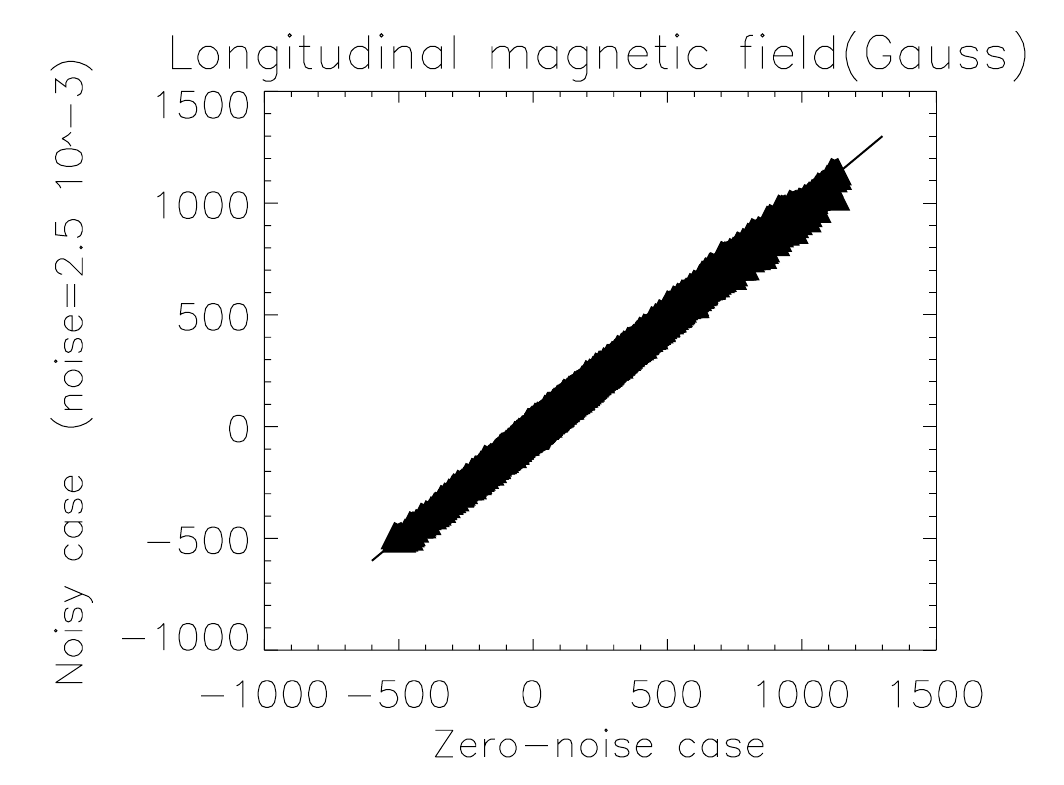}
 \includegraphics[width=7 cm]{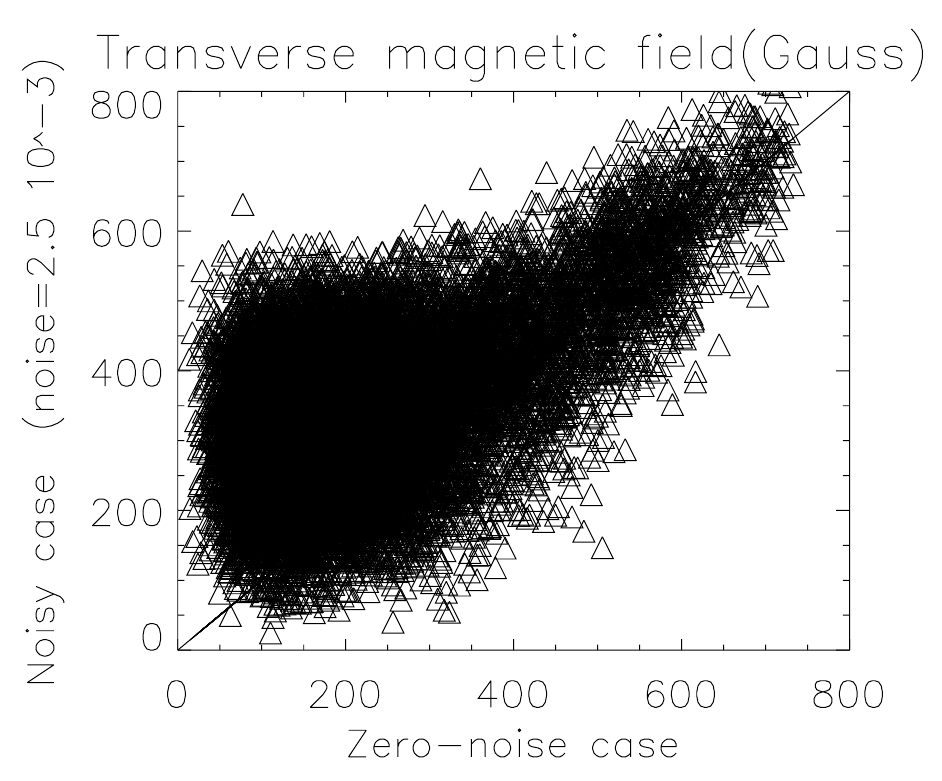}
\end{center}
\caption{ \textsf{
The same as Figure    \ref{figureB_noise10_4}    but the   noise level= 2.5 $\times$ 10$^{-3}$. }}
\label{figureB_noise2510_3}
\end{figure}

Let us now calculate the   error percentage on the magnetic field determination in the case of a  noise level=2.5 $\times$ 10$^{-3}$.    Figure   \ref{figureB_noise2510_3}  represents, with a solid line, the theoretical  exact solution   corresponding to the equation $y=x$ in the Cartesian coordinates.  Moreover, in the same figure,    the  open triangles   represent   the magnetic field obtained in the  noisy case as a function of the magnetic field derived in the zero noise case.  The difference between the noisy and non-noisy cases is less than 7 \% in the case of the longitudinal magnetic field, however, it can be up to 75 \% in the case of the transverse magnetic field. Thus,   for many pixels the noise is so important that it is impossible to derive the  transverse magnetic field from the observation of the linear polarization in the sodium line. It is worth mentioning that     the modest effective Land\'e g-factor of the Na I $\lambda$ 5896 \AA\ line  (g= 1.33)  contributes to an expected   weak linear polarization signal.    

As a conclusion, one should consider that  for a   noise level $>$   10$^{-3}$, it is difficult to   correctly interpret  the 
Na I linear polarization in terms of transverse magnetic field.   The situation
 must be better for the   Fe I line $\lambda$ 6302.5 \AA\ owing to its high sensitivity to magnetic fields. In fact,  its Land\'e factor (g = 2.5) is almost two times larger than the Land\'e factor of the Na I line. In any case, by inverting the synthetized Stokes  profiles of the Na I   line, our aim  was only to test our inversion method, and not to know anything about the solar magnetic field.

As an application, our numerical code is applied to  the Fe I  $\lambda$ 6302.5 \AA\  line which is widely used for solar magnetometry.
    
\section{MERLIN inversion code vs. our numerical code }
  We use   full spectra data
 of Hinode Solar Optical Telescope-Spectropolarimeter (SOT-SP) obtained for the Fe 6302.5 $\AA$  line  (Kosugi et al. 2007;  Tsuneta et al. 
2008; Lites et al. 2013). Stokes profiles (level 1) and level 2 outputs from inversions using the 
HAO  MERLIN  inversion code developed under the Community Spectropolarimetric Analysis Center are available online.

 This allows us to compare our code with the MERLIN code. It is worthy noticed  that  MERLIN inversions are 
performed at every pixel regardless of polarization profile, and the inversion code caps the field strength values at 5000 Gauss, so it  will not  give values larger than that. If a pixel reaches a value of 5000 Gauss, then one can assume that the code has not converged properly for that pixel. For other quiet Sun pixels, the magnetic field is about 3000 or 4000 Gauss but we verified that the corresponding polarization signals are weak which means that MERLIN gives clearly incorrect results for that pixels. Before the comparison, I removed the polarization profiles giving    spurious 
and unphysical results.


 Level 2 file chosen here is '20160903\underline{\quad}074908.fits' which corresponds to the inversion of 1139 profiles of polarization presented in level 1 data;   i.e. the slit of the SP instrument scanned the solar surface in 1139 steps to construct a 2D image   of the solar surface.  The level 1 data are calibrated profiles of polarization containing 3D data (spectral x spatial x 4 Stokes parameters)   ready for scientific analysis. 
 For direct comparison, I inverted the same level 1  profiles and confronted our results to the MERLIN's results.

\begin{figure}
\begin{center}
\includegraphics[width=14 cm]{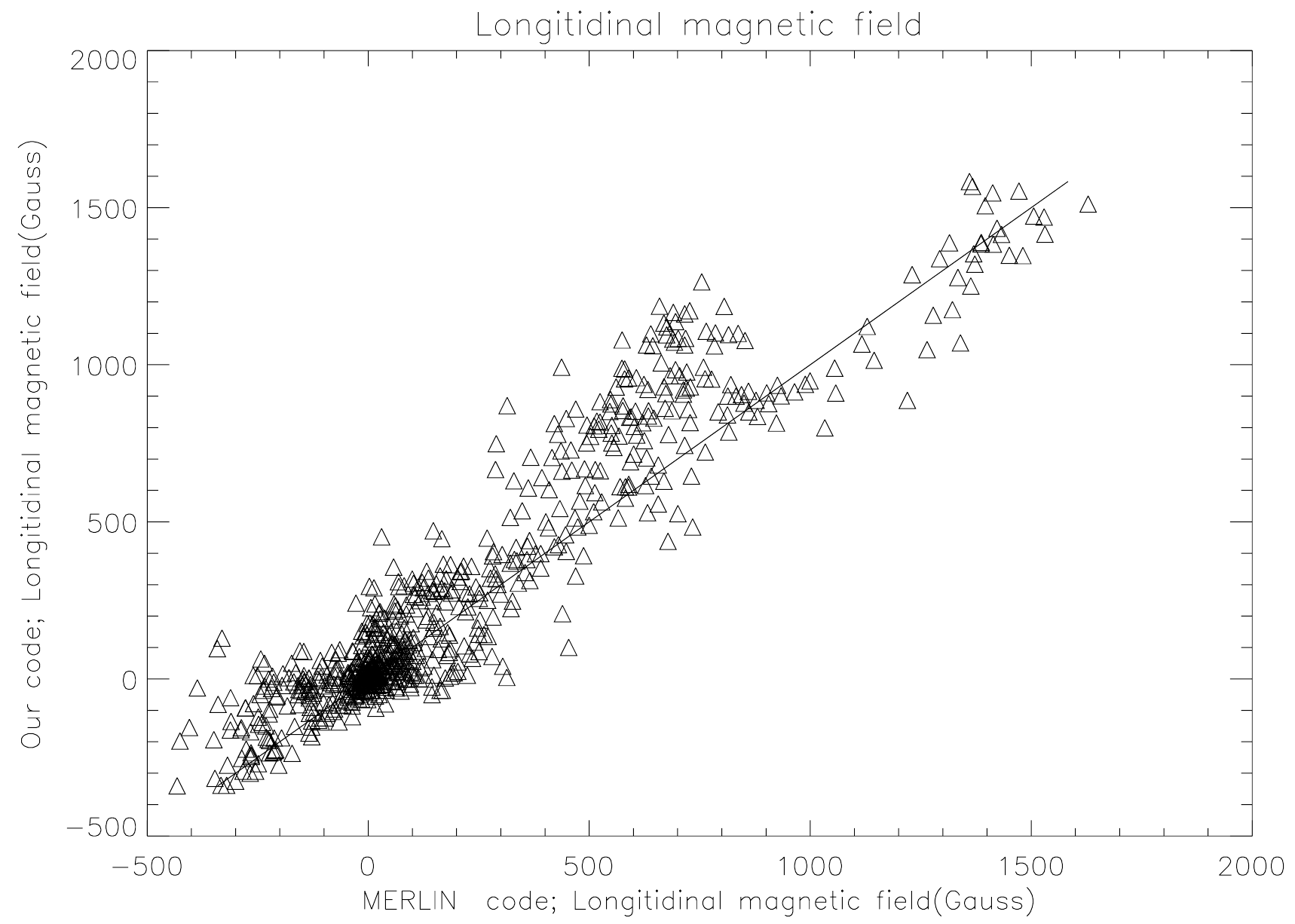}
\end{center}
\caption{ \textsf{
MERLIN inversion code VS our code.   }}
\label{RESPONSE_REFEREE}
\end{figure}

 The results are  encouraging and give us the conviction that the code provides sufficiently
 precise output  although that it is based on rather simple assumptions. 
  Figure    \ref{RESPONSE_REFEREE}    shows, with a solid line, the theoretical  exact solution corresponding to the equation of a straight line passing through the origin  and the result of the  comparison between the solution corresponding to our results and  the solution obtained by MERLIN.

\section{Application: inversion of the observations of the polarization profiles of the Fe {\sc i} $\lambda$ 6302.5 \AA\  line}
\subsection{Observations}
  \begin{figure}
\includegraphics[width=6.5  cm]{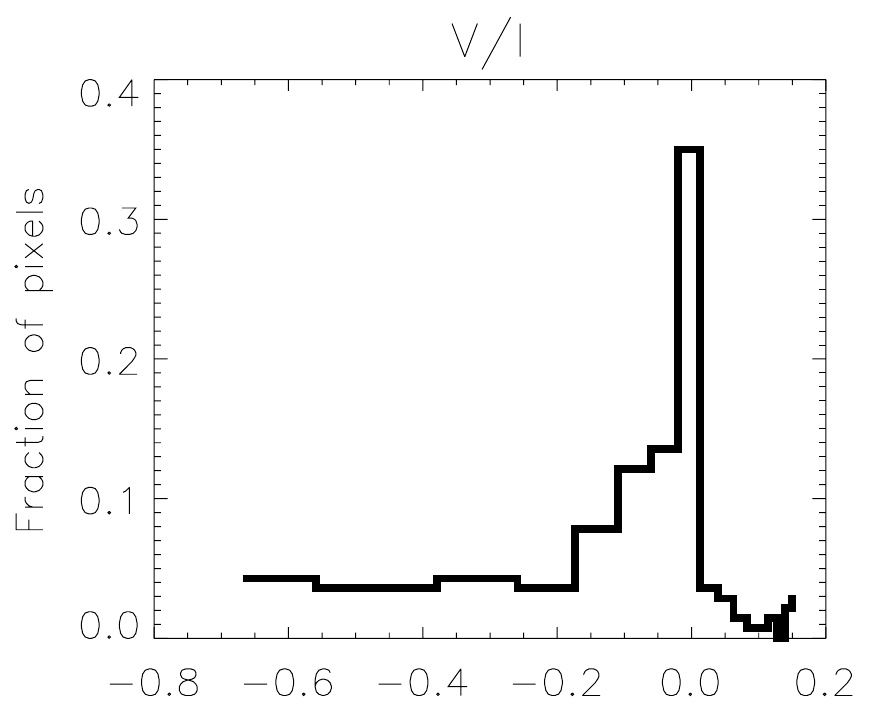}  
\includegraphics[width=6.5  cm]{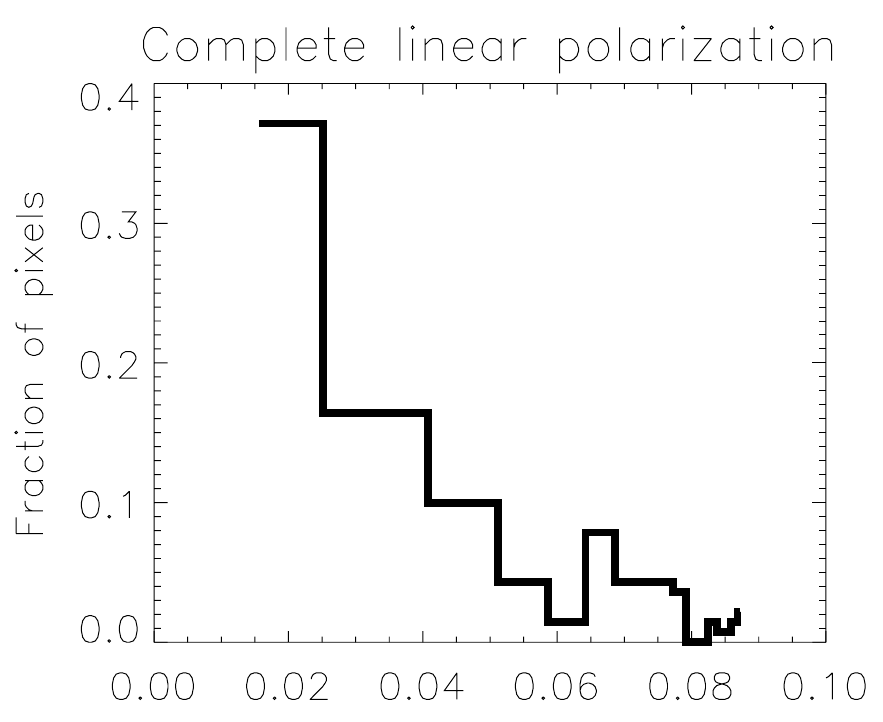}
 \caption{ \textsf{ Fraction of pixels at each range of values of the     circular polarization $V/I$ and complete linear polarization $\sqrt{Q^2+U^2}/I$ which are directly related to the longitudinal and transverse magnetic field, respectively.}}
\label{figureVandPL}
\end{figure}

   \begin{figure}
\includegraphics[width=  6.5 cm]{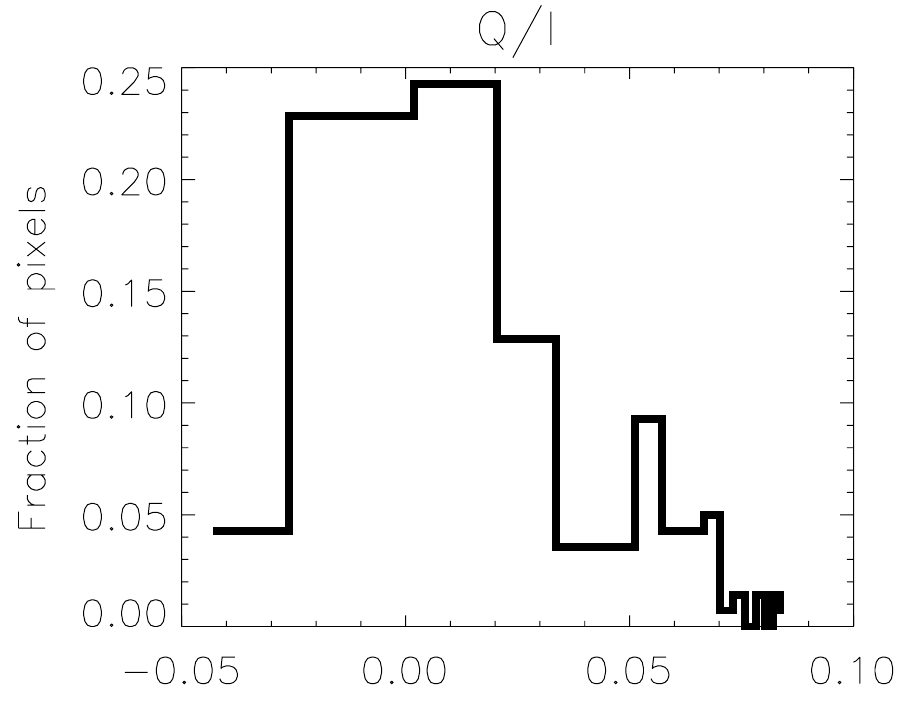}
\includegraphics[width= 6.5 cm]{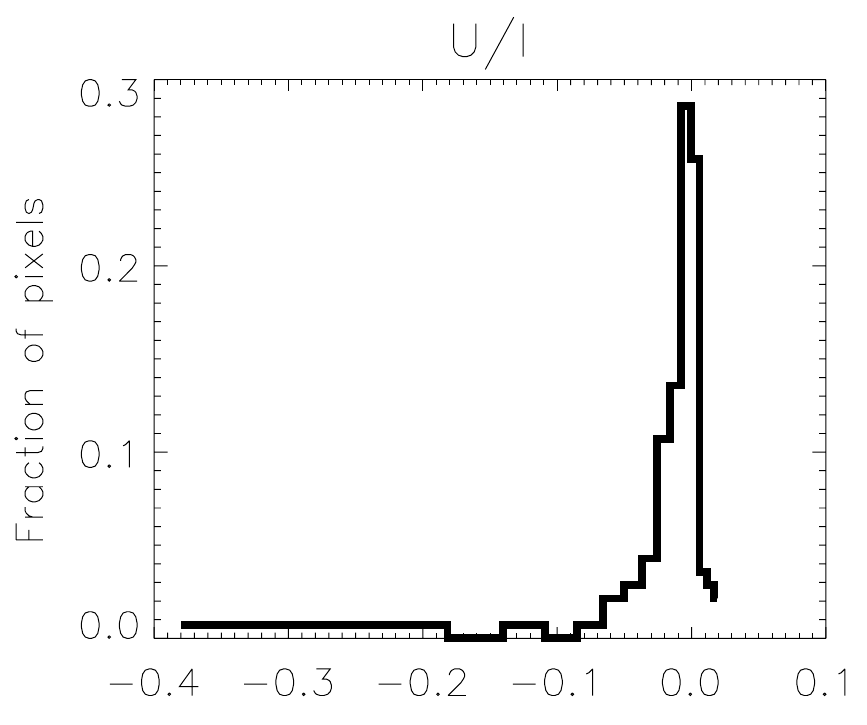}
\caption{ \textsf{Fraction of pixels at each range of values of the  Stokes parameters  $Q/I$ and $U/I$.}}
\label{figureQandU}
\end{figure}

The observations of the polarization profiles of the Fe {\sc i} $\lambda$ 6302.5 \AA\  line has been kindly communicated to us by
 Dr.  Michele Bianda (IRSOL). The   observations were performed on April 10, 2016, at IRSOL in Locarno using the 45 cm aperture Gegory Coud\'e telescope, the 10 m focal length Czerny-Turner spectrograph (grating 180 mm x 360 mm, 316 lines / mm), and ZIMPOL (Ramelli et al. 2010).
 
The observed sunspot was AR2529 located near the East limb. The spectrograph slit,  oriented parallel to the solar polar limbs, or perpendicular to the solar rotation axis, was crossing the sunspot. The slit-width of 60 $\mu$ corresponds on the solar image to ~0.5   arcsec.
 
The high modulation rate of 1 kHz delivered by the FLC modulator of ZIMPOL permits to overcome   spurious polarization signatures originated by seeing effects. The FLC modulator is designed following the Gisler method (Gisler 2005).
 
Reduced data  is the combination of 4 CCD recordings of 0.2 $s$ exposure each (see Ramelli et al. 2010 for details). That is a short time compared to the usual ZIMPOL observations intended to measure scattering polarization signatures where precision in the order down to  $10^{-5}$ is required. In the observation reported here  a noise in the order of $2 \times$ $10^{-3}$ is reached, but the short exposure reduces the image quality degradation originated by the seeing. Consequently, as one is working with large signatures, that is an advantage.

The   observing angle is $\sim$ $50^{o}$.  For this observation angle,   the complete linear polarization $\sqrt{Q^2+U^2}/I$ reaches the  values    larger than the circular polarization $V/I$. The transverse component of the magnetic field is   expected to be larger than the longitudinal one.  It is worth mentioning that if   the angle of observation is changed, the observed Stokes profiles  will be changed. Thus,    the  components $\vec{B}_{\small Trans}$  and $\vec{B}_{\small LOS}$ will be changed. However, the magnetic field vector $\vec{B}$= $\vec{B}_{\small Trans}$  + $\vec{B}_{\small LOS}$ does not depend on the angle of observation.
The inversion provides a unique  output (i.e. unique magnetic field vector).   

We notice in the Figures  \ref{figureVandPL}  and   \ref{figureQandU}  that about   30\% of the observed pixels have a  small polarization. That will directly affect the values of the magnetic field.
\subsection{Determination of the magnetic field vector}
 \begin{figure}
\begin{center}
\includegraphics[width=12 cm]{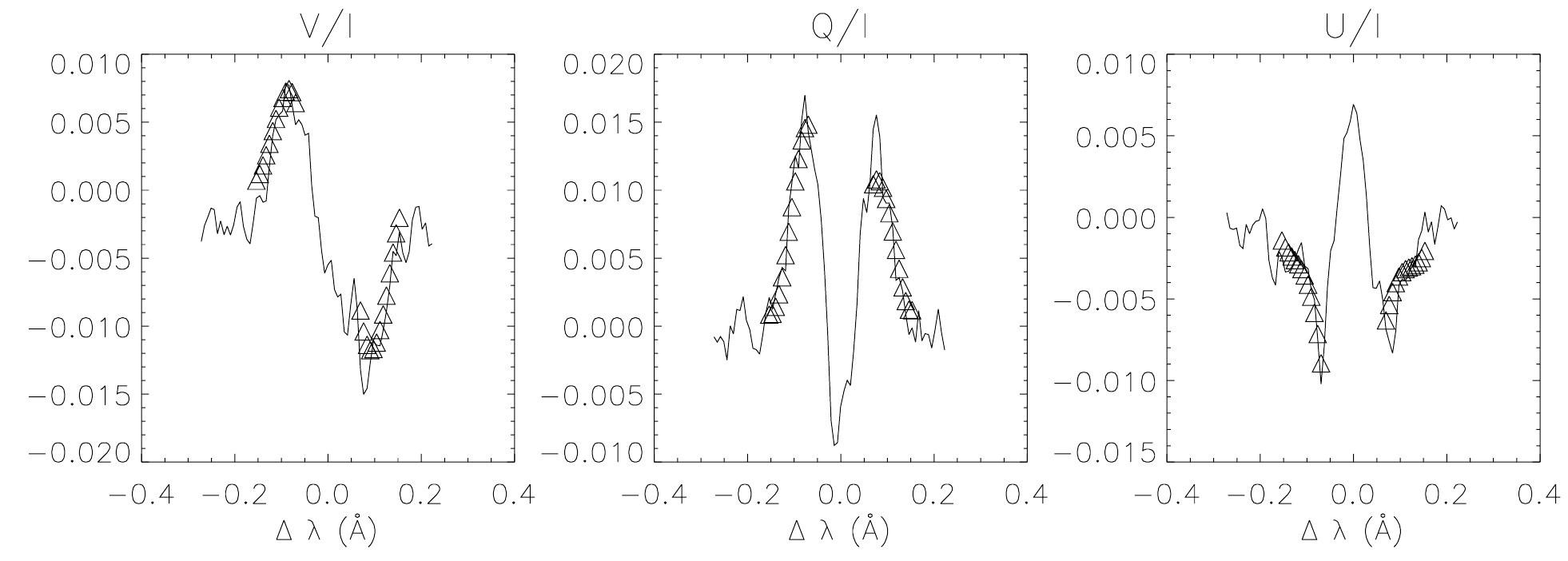}
\end{center}
\caption{ \textsf{ Example of the fitting of the observations using SVDC and SVSOL methods  for a   Q, U and V parameters. The tolerance adopted is $\delta\lambda_{TOL}$=0.045 \AA  and the place of the  target wavelength	  $\delta\lambda_B$= 0.109863 \AA.   The  open triangles ($\triangle$)  represent  the theoretical profile  resulted from the fitting procedure.   }}
\label{Stokes-Fit}
\end{figure}

\begin{figure}
\includegraphics[width=  6.5 cm]{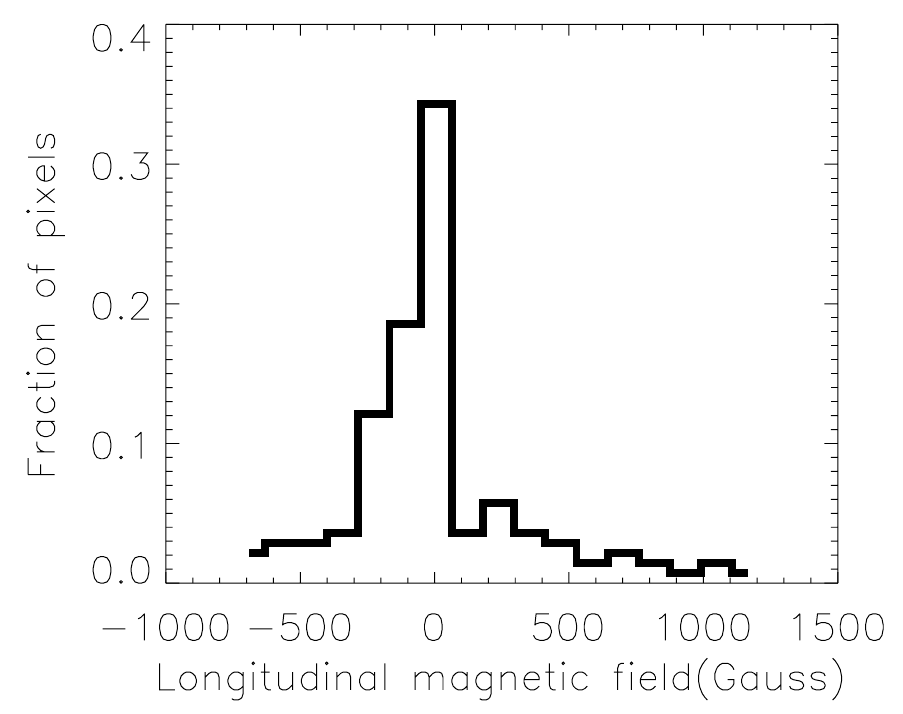}
\includegraphics[width= 6.5 cm]{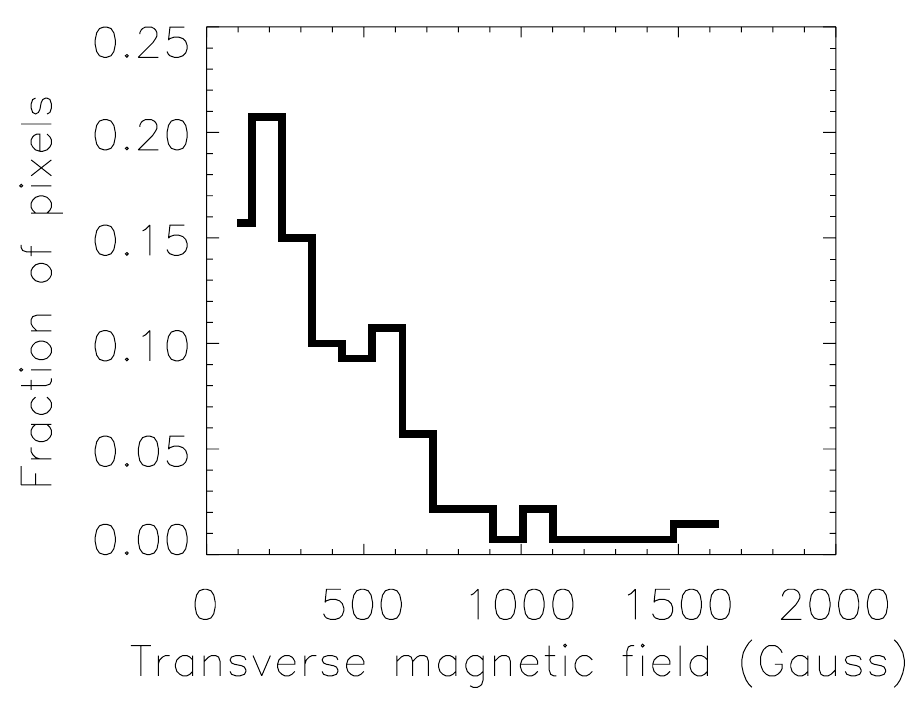}
\caption{ \textsf{Fraction of pixels at each range of values of the   transverse and  the longitudinal components of the magnetic field.}}
\label{figureBlongBtrans}
\end{figure}

We started by fitting the observed intensity and polarization profiles. An example of the results of the fit is presented in
 the Figure  \ref{Stokes-Fit}. Then we  computed the magnetic field vector inferred from the observations.  
The results are presented in histograms that count  the fraction of pixels in each range of values. The majority of the observed pixels have a small value of the magnetic field. There is a clear correlation between the values of the magnetic fields presented in the Figure  \ref{figureBlongBtrans}  and those of the polarization in the Figures  \ref{figureVandPL}  and   \ref{figureQandU}. 

\begin{figure}
\includegraphics[width=6.5  cm]{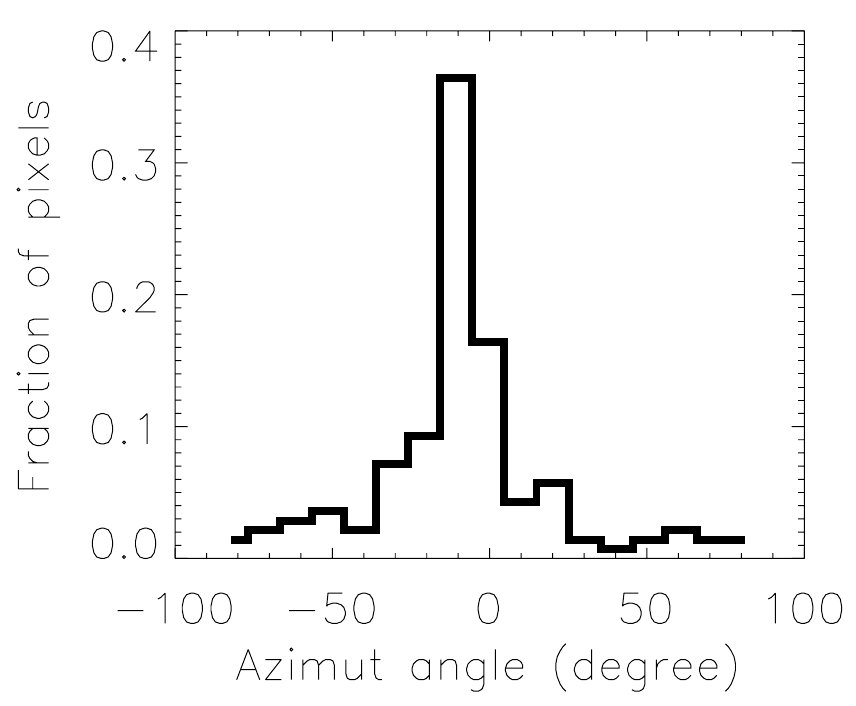}  
\includegraphics[width=6.5  cm]{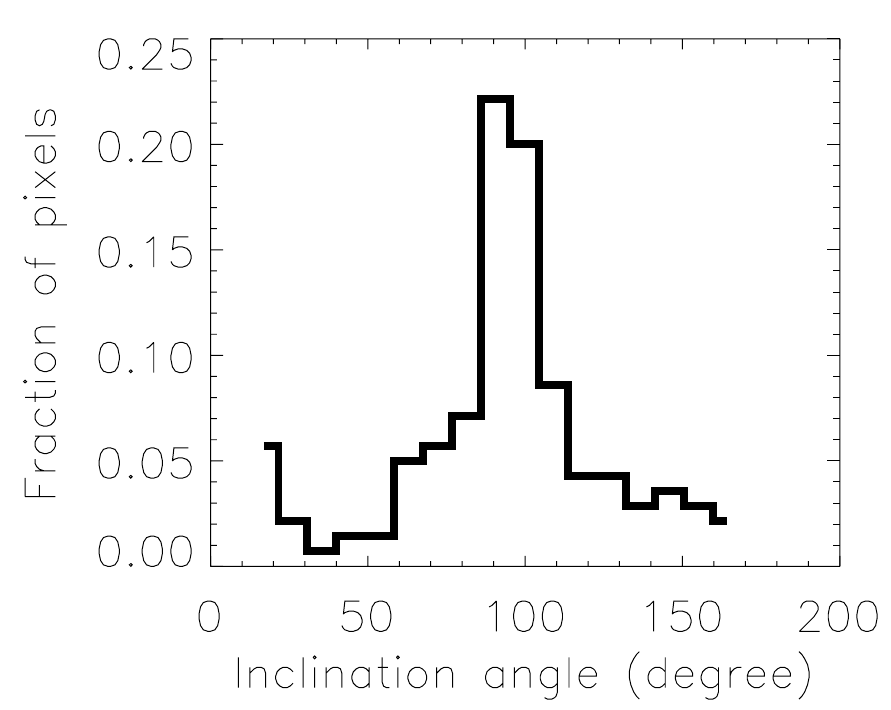}
 \caption{ \textsf{Fraction of pixels at each range of values of the  azimuthand  the inclination angles.}}
\label{figureAzimutINCLIN}
\end{figure}

Figure  \ref{figureAzimutINCLIN}  indicates that  between $\sim$ 10 \% and 20 \% of pixels have a magnetic 
field inclined of about  80$^{\textrm{o}}$ to 100$^{\textrm{o}}$ from the horizontal. The azimuth angle is mainly around 10$^{\textrm{o}}$.

\section{Final remark: combining the bisector method with our inversion method}
The bisector method tells us that:
\begin{eqnarray}\label{eq_Bbisector1}
B_{\small LOS}(Gauss)=\frac{1.071 \times 10^{12}}{g \times \lambda_0^2} 
\times  (\lambda_+-\lambda_-)
\end{eqnarray}
where $\lambda_+$ is the central wavelength or bisector location of $I+V$ 
profiles  and $\lambda_-$  is the bisector location of $I-V$ profiles 
(Rayrole 1967, Semel 1967). 

By comparing $B_{\small LOS}$ of the Equation  \ref{eq_BLOS}    and the 
Bisector expression of $B_{\small LOS}$   given by  Equation  \ref{eq_Bbisector1}, one concludes 
that both are exactly the same if:
\begin{eqnarray}\label{eq_Bbisector-JL{Introduction}HJS}
\frac{-2 \times V}{(\frac{\partial I}{\partial \lambda})} = (\lambda_+-\lambda_-)
\end{eqnarray}
The value of $\frac{\partial I}{\partial \lambda}$ can be determined using the bisector method and then
introduced in the Equation  \ref{eq_BTrans}. Thus the bisector method could be used to obtain the LOS magnetic field but also could contribute in the determination of the transverse magnetic field. It is worth noticing that the bisector method is  valid even for strong magnetic fields.  

\section{Conclusions}
Zeeman polarization in the solar  lines  can be a crucial source of information about   
photospheric and chromospheric magnetic fields.

 The main concern of our work was to develop a new inversion code based on Zeeman effect and to evaluate its accuracy for future applications. We showed that our code gives results with a satisfactory precision.  For a given input magnetic field configuration,  one can  synthesize   
  Stokes profiles emergent from solar  atmosphere  for any value of the observation angle    $\mu$.  
  We inverted the data  obtained   for  $ \mu$=1. Using the method presented in our work, it is possible to determine the variation of the magnetic field vs the target wavelength $ \delta \lambda_B$ which is equivalent to the variation of the magnetic field with the height in the atmosphere. The choice of the tolerance follows the best compromise that we found between stability of the  inversion method and obtaining the magnetic variation as a function of the wavelengths.    As an additional test to the precision of the inversion code,  we used   the level 1 and level 2 data available online at the Community Spectro-polarimtetric Analysis Center  to compare our results with the results obtained with the HAO MERLIN inversion code.  

Finally, as an application, we inverted real observations that   were performed on April 10, 2016, at IRSOL in Locarno.

\section*{Acknowledgments}
I am greatly indebted to Dr. Michele Bianda (IRSOL, Switzerland)   for communicating to me   unpublished observations and I would like to thank  Dr. Graham Barnes and Dr. K.D. Leka for 
  helpful discussions during  my stay in  CoRA (USA).  HINODE is a Japanese mission developed and launched by ISAS/JAXA, with NAOJ as domestic partner and NASA and STFC (UK) as international partners. It is operated by 
these agencies in cooperation with ESA and NSC (Norway). Hinode SOT/SP 
Inversions were conducted at NCAR under the framework of the Community 
Spectro-polarimtetric Analysis Center (CSAC; http://www.csac.hao.ucar.edu).

\end{document}